\documentclass[letterpaper,11pt]{article}%\usepackage{latexsym,fullpage,multirow,ltexpprt}
\begin{document}
\addtolength{\textheight}{+0.18in}
 
\newtheorem{claim}{Claim}[section]
\newtheorem{lemma}{Lemma}[section]
\newtheorem{theorem}{Theorem}
\newtheorem{cor}{Corollary}[section]
\newtheorem{prop}{Proposition}[section]
\newcommand{\lf}{\left\lfloor}
\newcommand{\rf}{\right\rfloor}
\newcommand{\dis}{\displaystyle}
\newcommand{\sums}[2]{\sum_{\stackrel{{\scriptstyle {#1}}}{{#2}}}}
\newcommand{\prods}[2]{\prod_{\stackrel{{\scriptstyle {#1}}}{{#2}}}}
\newtheorem{Definition}{Definition}
\newtheorem{corollary}{{Corollary}}
\newcommand{\beq}{\begin{eqnarray*}}
\newcommand{\eeq}{\end{eqnarray*}}
\newcommand{\letab}{\le &}

\newenvironment{tabAlgorithm}[2]{
\setcounter{algorithmLine}{1} \samepage
\begin{tabbing}
999\=\kill #1 \ \ --- \ \ \parbox{3.8in}{\it #2} }{
\end{tabbing}
}
\newcounter{algorithmLine}
\newcommand{\algline}{\\\thealgorithmLine\hfil\>\stepcounter{algorithmLine}}
\newcommand{\algnono}{\\ \>}

\newcommand{\TRUE}{{\bf TRUE}}
\newcommand{\FALSE}{{\bf FALSE}}
\newcommand{\NIL}{{\bf NIL}}
\newcommand{\CURRENT}{{\bf CURRENT}}
\newcommand{\IF}{{\bf IF }\=}
\newcommand{\THEN}{{\bf THEN }\=}
\newcommand{\ELSE}{{\bf ELSE }}
\newcommand{\WHILE}{{\bf WHILE }\=}
\newcommand{\FOR}{{\bf FOR }\=}
\newcommand{\DO}{{\bf DO }\=}
\newcommand{\RETURN}{{\bf RETURN }}
\newcommand{\BREAK}{{\bf BREAK }}

\input epsf
 
\def\prbox{\hfill\rule{1.2ex}{1.2ex}\vspace{.2in}}
 
\newenvironment{proof}{\noindent{\bf Proof }}{\prbox}
 
\def\Reals{\hbox{\rm I\kern-.18em R}}
\def\reals{\Reals}
\def\Complexes{\hbox{\rm C\kern-.43em
       \vrule depth 0ex height 1.4ex width .05em\kern.41em}}
\def\complexes{\Complexes}
 
\def\field{\hbox{\rm I\kern-.18em F}} %symbol for field
\def\Naturals{\hbox{\rm I\kern-.17em N}}
\def\naturals{\Naturals}
\def\integers{\hbox{\rm Z\kern-.3em Z}}
\def\hh{\hrule height0.9pt width1.1em}
\def\vv{\vrule width0.8pt depth0.12em height0.92em}
\def\square{\vbox{\kern0.15em\hh\kern0.9em\hh\kern-1.05em
            \hbox{\vv\kern0.93em\vv}}}
\def\Lv{L_{\alpha} }
\def\Lvp{L_{\alpha'} }
\def\sm{{\mbox{\boldmath $\sigma$}_m}}
\def\MMv{R_{\mbox{\boldmath $\sigma$}_v}}
\def\tMMv{\widetilde{R}_{\mbox{\boldmath $\sigma$}_v}}
\def\MMm{R_{\mbox{\boldmath $\sigma$}_m}}
\def\tMMm{\widetilde{R}_{\mbox{\boldmath $\sigma$}_m}}
\def\MMj{R_{\mbox{\boldmath $\sigma$}_j}}
\def\sn1{{\mbox{\boldmath $\sigma$}_{n-1}}}
\def\snv{{\mbox{\boldmath $\sigma$}_v}}
\def\eps{\epsilon}
\def\D{\Delta}
\def\tb{\hspace*{0.5 in}}
%\hyphenation{eigen-value so-lu-tions heur-is-tics mod-eled de-note
%prob-a-bil-ity start-ing}

\title{Approximation Algorithms for the Bipartite Multi-cut problem}

\author{%
\begin{tabular}{p{5cm}p{5cm}}
  \centering Sreyash Kenkre & Sundar Vishwanathan \\
  \multicolumn{2}{c}{\small Department Of Computer Science \& Engineering,} \\
  \multicolumn{2}{c}{\small IIT Bombay, Powai-400076, India.}\\
  \multicolumn{2}{c}{\small {\tt \{srek,sundar\}@cse.iitb.ac.in}}
\end{tabular}%
}

\date{}
\maketitle

\begin{abstract}
We introduce the {\it Bipartite Multi-cut} problem. This is a generalization of the {\it st-Min-cut} problem,
is similar to the {\it Multi-cut} problem (except for more stringent requirements) and also turns out to be
an immediate generalization of the {\it Min UnCut} problem. We prove that this problem
is {\bf NP}-hard and then present 
LP and SDP based approximation algorithms. While the LP algorithm is based on the Garg-Vazirani-Yannakakis
algorithm for {\it Multi-cut}, the SDP algorithm uses the {\it Structure Theorem} of $\ell_2^2$ Metrics.
\end{abstract}

\section{Introduction}
Given a graph $G=(V,E)$ with non negative weights on its edges, the {\it st-Min-cut} problem asks for the minimum 
weight subset of edges, whose deletion disconnects two specified vertices $s$ and $t$. This is a well studied
problem and can be solved polynomial time. However there are many generalizations of this, like the 
{\it Multiway Cut} and the {\it Multi-cut}, which are $\mathbf{NP}$-complete. We introduce one such problem, which we call
as the {\it Bipartite Multi-cut} or $BMC$.\\

\noindent {\bf Problem:} Bipartite Multi-cut\\
\noindent {\bf Input:} Graph $G=(V,E)$, non negative weights $w_e$ on every edge $e\in E$,
 $k$ {\it source-sink} pairs $(s_1,t_1),\ldots,(s_k,t_k)$.\\
\noindent {\bf Output:} $X$ and $\overline{X}$ such that $|\{s_i,t_i\}\cup X|=1$, $\forall i=1,\ldots,k$.\\

When $k$ is one this is the usual {\it st-Min Cut} problem.
We show that it is a natural and immediate generalization of the {\it Min UnCut} problem and hence is {\bf NP}-hard.
We first show a {\it linear programming} (LP) based $O(\log k)$ factor approximation algorithm. 
The LP for multi-cut~\cite{GVY,VV} is also a relaxation for {\it BMC} but it can be seen that its integrality is
$\Omega(k)$ for {\it BMC}. We need to add some {\it symmetrization} constraints to obtain the $O(\log k)$
approximation ratio.
We then improve the 
approximation factor to $O(\sqrt{\log k}\log\log k)$ using {\it semidefinite programming} (SDP). 
This is in contrast to the {\it Multi Cut} problem, 
where no improvement to the $O(\log k)$ factor approximation has as yet been reported.
The SDP for {\it BMC} is similar to that of {\it Min Uncut}~\cite{ACMM}.
However their weighted separation techniques are not applicable to our problem as we do not have 
enough symmetry. We need a stronger analysis combining the region growing
algorithm for {\it Multi Cut} by Garg, Vazirani and Yannakakis~\cite{GVY} and the {\it structure theorem} for
$\ell_2^2$ metric spaces due to Arora, Rao and Vazirani~\cite{ARV}.

In the next section we prove hardness results. We then present the LP and SDP relaxations and then review the
analysis of the region growing techniques. We follow this with a description of the algorithms and prove
the approximation guarantees.

\section{Hardness}
To prove hardness we reduce the {\it Min UnCut} problem to {\it BMC} in an approximation preserving way. 

\begin{Definition}(Min UnCut)\\
Give boolean constraints of the form $x_i \oplus x_j = 0$ and $x_i \oplus x_j = 1$, where $x_1,x_2,\ldots,x_n$ are
boolean variables, find an assignment that minimizes the number of unsatisfied constraints.
\end{Definition}

We use a construction from~\cite{ACMM} to reduce this problem to {\it BMC}. Consider a graph on $2n$ vertices
$\{v_1,v_2,\ldots,v_n\}\cup\{v_{-1},v_{-2},\ldots,v_{-n}\}$. A variable $x_i$ and its complement correspond to the
vertices $v_n$ and $v_{-n}$ respectively. For each constraint of the form $x_i\oplus x_j = 0$, put an edge between
$v_i$ and $v_j$ and between $v_{-i}$ and $v_{-j}$. For each constraint of the form $x_i\oplus x_j = 1$, put an edge between
$v_i$ and $v_{-j}$ and between $v_{-i}$ and $v_j$. Give all the edges unit weight. The {\it Min UnCut} problem is the same
as that of finding a partition $X,\overline{X}$ of $V$ that separates the pairs $(v_1,v_{-1}),\ldots,(v_n, v_{-n})$
and minimizes the number of edges crossing the cut (this gives twice the cost).
This proves that {\it BMC} is {\bf NP}-hard.

By trying all assignments of $s_i$ and $t_i$ to $X$ and $\overline{X}$, {\it BMC}
can be solved exactly by using $2^k$ flow computations, so that it is polynomial time solvable if
$k=O(\log n)$. Further, we show below that it suffices to only consider the case when each source or sink occurs
in exactly one pair. Consider a graph $D$ (called as the {\it demands graph}) with vertex set as
$\{s_1,s_2,\ldots,s_k\}\cup\{t_1,t_2,\ldots,t_k\}$, with
an edge between every source and its corresponding sink (some sources may have multiple sinks). For {\it BMC} to be
feasible, $D$ should be bipartite. Consider the connected
components of this bipartite graph. If we fuse the vertices in $G$, corresponding to each side of a component of the
demands graph and solve the problem on this fused graph, the solution is easily seen to be equal to that of the the
original problem. Hence from now on we assume that the demands graph is a matching.

Every solution to the {\it BMC} problem is a feasible solution to the {\it Multi Cut} problem. However unlike the
{\it Multi Cut} problem, we prove below that even on paths {\it BMC} is as hard as the general case
(see appendix for an $O(n^2)$ exact algorithm for multi cut on paths).

Consider any instance of {\it BMC} on a graph $G$.
Since the number of odd degree vertices in a graph is even, by putting a dummy vertex and connecting it
to all the odd degree vertices using zero weight edges we can make all degrees even, without affecting the solution.
Hence we assume that the given instance is on an Eulerian graph~\cite{west}.
Let $v_1,e_1,v_2,e_2,\ldots,e_m,v_1$ be an Euler circuit and let $P$ be the
path obtained by laying out this circuit linearly. We denote by $v_i^1, v_i^2,\ldots,v_i^{d_G(v_i)}$ the different
copies of $v_i$ occurring in $P$. Let $P'$ be the path $u_1,e'_1,u_2,e'_2\ldots,u_{n-2k}$, where each $u_i$ corresponds to
a vertex in $G$ which is neither a source nor sink. We give zero weights to the edges $e'_i$ of $P'$. Consider the path
$P''$ obtained by joining vertex $v_1$ of $P$ to vertex $u_1$ of $P'$ using a zero weight edge.
We define a $BMC$ instance on $P''$. The source-sink pairs are $(v_{s_i}^x,v_{t_i}^y)$,
for all $x=1,\ldots,d_G(s_i)$ and $y=1,\ldots,d_G(t_i)$, where $d_G(v)$ denotes the degree of vertex $v$ in $G$.
Also, for every vertex $v$ in $G$ that is neither a source nor a sink, if $u_i$ is the corresponding vertex in $P'$,
then we add $(u_i,v^1),\ldots,(u_i,v^{d_G(v)})$ as source-sink pairs.
Let $X,\overline{X}$ be a feasible solution for $P''$. Let $Y$ and $\overline{Y}$ be the vertices in $G$
corresponding to the vertices in $X\cap P$ and $\overline{X}\cap P$ respectively. 
Since for all $x=1,\ldots,d_G(s_i)$, $(v_{s_i}^x,v_{t_i}^1)$ are source sink pairs, all $v_{s_i}^x$ lie in either
$X$ or $\overline{X}$. Similarly, all $v_{t_i}^y$ lie in either $\overline{X}$ or $X$ and all copies of a non source-sink
vertex lie either in $X$ or $\overline{X}$.
Using this it is easily seen that $Y$ and $\overline{Y}$ is a feasible solution for $G$, with the same cost as
$X$,$\overline{X}$. Conversely, by a similar argument, every feasible
solution to $G$ corresponds to a feasible solution of $P''$ with the same cost. Thus {\it BMC} restricted to paths is
as difficult as the general case.

\section{LP and SDP Relaxations}
Let $\mathcal{P}$ denote the set of paths that connect some $s_i$ to the corresponding $t_i$. For vertices $v_i$ and $v_j$
we associate a distance $d_{v_iv_j}$. If $v_iv_j=e\in E$, then we refer to $d_{v_iv_j}$ as $d_e$.
Consider the following linear program.

\begin{eqnarray}\label{LP} 
\hbox{ min }\sum_{e\in E}w_e d_e \;\rule{1cm}{0pt}
\end{eqnarray}
\begin{eqnarray}\label{LP} 
\sum_{e \in P} d_e & \geq & 1,\;\;\forall P \in \mathcal{P} \label{everypath} \\
d_{v_iv_j} + d_{v_jv_k} & \geq & d_{v_iv_k}\hbox{,  }\forall v_i,v_j,v_k \in V
\end{eqnarray}
\begin{eqnarray}\left.\begin{array}{rrr}
\rule{2cm}{0pt}\;d_{s_it_j} & = & d_{t_is_j}  \\
d_{s_is_j} & = & d_{t_it_j} 
\end{array}\right\rgroup \hbox{ }\forall i,j=1,\ldots,k\label{symmetry}
\end{eqnarray}
\begin{eqnarray}
d_{v_iv_j} & \geq & 0 \rule{2cm}{0pt}  \label{nonnegative}
\end{eqnarray}

Though this LP can have an exponential number of constraints, it can be solved because of a polynomial time
oracle to check feasibility- a shortest path procedure to check (\ref{everypath}), while the rest of the inequalities can
be checked in polynomial time. 
Suppose $X$ and $\overline{X}$ is an optimal integral solution for an instance of the problem.
For all $v\in X$ and $u\in \overline{X}$, set
$d_{vu}=1$. Set all other distances to zero.
Then $d_e=0$ for edges that have both end points in one of $X$ or $\overline{X}$, and $d_e=1$ otherwise.
It can be checked that this is feasible for the above LP. 
We prove that this LP can be rounded to give
a $O(\log k)$ factor algorithm. Note that the objective function, along with the inequalities~(\ref{everypath})
and (~\ref{nonnegative}) give the LP for {\it Multi Cut} used in~\cite{GVY}. However, unlike the {\it Multi Cut} where
no improvement on the LP based algorithm is known, we can improve the approximation guarantee by using SDPs.

We use the following SPD to give an $O(\sqrt{\log k}\log\log k)$ factor approximation algorithm.

\begin{eqnarray}\label{SDP}
\hbox{ min }\frac{1}{4}\sum_{e=uv\in E}w_e |x_u-x_v|^2  \\
|x_u -x_v|^2 + |x_v-x_w|^2 & \geq & |x_u-x_w|^2  \hbox{  } \forall u,v,w \in V \nonumber \\
|x_{s_i} - x_{t_i}|^2 &= & 4 \hbox{  } \forall  s_i, t_i \label{antipodal}\\
|x_v|^2  &= & 1 \hbox{  } \forall v\in V, \hbox{ }x_v\in \Re^n \nonumber
\end{eqnarray}

This SDP is same as the SDP for {\it Min Uncut}~\cite{ACMM}, except for the possibility that not all
vertices are sources or sinks.

Let $X$ and $\overline{X}$ be an optimal solution to this problem. Assign any unit vector $\mathbf{e}\in \Re^n$ to points 
in $X$ and $-\mathbf{e}$ to points in $\overline{X}$. Then it is easily seen that this assignment obeys the above
inequalities. The solutions to the SDP give an $\ell_2^2$-metric space on $V$~\cite{ARV}. It is also well known that the 
lengths $d_e$ assigned to the edges by a solution of the LP gives rise to a metric space on $V$~\cite{VV}~\cite{GVY}.
Without loss of generality, we assume that the graph $G$ is a complete graph with zero weight on those edges whose 
weights are not specified. Let $d_{uv}$ be a metric defined on $V$ (whether it is obtained by solving the LP or the SDP
shall be clear from the context). Let $V^*$ be the total {\it volume} of the metric space.

\begin{equation}
V^* = \sum_{u,v\in V}w_{uv}d_{uv}
\end{equation}

This is the value returned by the LP or the SDP, depending on which is used for defining the metric on $V$.
For a set $S$ and a vertex $v$, let $dist(v,S)$ denote the distance between $v$ and $S$, which is defined to be
$\hbox{min}_{u\in S} d_{vu}$.

\noindent For a vertex $v$, let $B(v,r)$ denote the subset of vertices in a ball of radius $r$ around $v$. 
That is $B(v,r) = \{ u\hbox{ }:\hbox{ } u\in V, d_{uv}\leq r\}$.
If $S=\{w_1,w_2,\ldots,w_{\ell}\}$ is a subset of $M$, let $B(S,r)=\cup_{w_i\in S}B(w_i,r)$. Define the volume of
$B(S,r)$, denoted by $V(S,r)$, which shall be shortened to $V(r)$ if the set $S$ being referred to is clear from context, as
follows ($V(0)$ denotes the initial volume on $S$, if any).

\begin{equation}
V(r)= \sum_{u,v\in S(r)}w_{uv}d_{uv} + \sum_{u\in S(r),v\not\in S(r)} w_{uv}d_{uv}\frac{r-dist(S,u)}{dist(S,v)-dist(S,u)}+V(0)
\end{equation}

\noindent Clearly, $V(\infty) = V^*$. 
Let $C(S,r)$, referred to as the {\it cut}, be the total weight of the edges crossing $B(S,r)$. When there is no 
possibility of confusion, we refer to this as $C(r)$.

\begin{equation}
C(r) = \sum_{u\in S(r),v\not\in S(r)} w_{uv}
\end{equation}

$M$ shall denote the submetric induced by the sources $s_i$ and sinks $t_i$. A subgraph of $G$ is said to be {\it symmetric}
if it contains a source $s_i$, if and only if it contains the sink $t_i$
(it may also contain any subset of the non-source and non-sink vertices).
Two subsets $A$ and $B$ are called {\it antipodal} if sinks of the sources, and sources of the sinks of $A$ are
contained in $B$, and {\it vice versa}.

\section{Region Growing}
Our algorithms depend on region growing to get a feasible solution.
It differs from the algorithm of Garg, Vazirani and Yannakakis~\cite{GVY} in two ways. First, we grow regions around
subsets of the vertex set and second, we grow regions simultaneously around two subsets.
In the algorithm the two subsets $X$ and $\overline{X}$
are constructed
simultaneously and iteratively. At each step two subsets of vertices $A$ and $B$ are chosen, and one of them is assigned 
to $X$ and the other to $\overline{X}$.
We need this to enforce the conditions that the sources and sinks lie in different parts.
This is done by requiring that $A$ and $B$
be so chosen that $G-A-B$ is symmetric. This means that if a source or sink is contained in $A$, the corresponding
sink or source should be contained in $B$, i.e., $A$ and $B$ are antipodal (similar to~\cite{ACMM}). 
However unlike their technique where charging
the cut to the total volume suffices to obtain a good approximation guarantee, we need to charge the cut to the volume
of the grown region, as well as the total volume, depending on the initial volume of the regions.
This is because, while for their algorithm $G$ was the same as $M$, the volume of $M$ in our setting
may be arbitrarily smaller then the volume of $G$. Note that
due to the antipodal constraints (equation~\ref{antipodal}) for the SDP, and the symmetry constraints
(equation~\ref{symmetry}) if we start the region growing from antipodal subsets, we get
antipodal subsets at the same radii. The existence of common radii that obey the cut to volume charging constraints
follows from the analysis of the next section.

\section{Random Cuts}
The purpose of region growing is to charge the resulting cut to either the total volume~\cite{ARV,ACMM} or the
volume enclosed~\cite{GVY}. Typically, one uses an averaging argument to prove that a certain cut has a small weight. 
In most cases the proofs also yield the stronger fact that a random cut works with high probability.

Suppose $(V,d)$ is a metric space on the vertex set $V$ of graph $G$. Let $d_{uv}$ denote the distance between
the vertices $u$ and $v$. Let $w_{uv}$ be the weight of the edge $uv$. Let $S=\{w_1,w_2,\ldots,w_{\ell}\}$ be a set 
of vertices which form the centers of expansion of the region. 

\begin{theorem}\label{average}
Let $ 0\leq r_1 < r_2$ be two radii. Then with probability at least $3/4$ a random radii between
$r_1$ and $r_2$ has a cut of size at most $\frac{4}{r_2-r_1}(V(r_2)-V(r_1))$.
\end{theorem}

Since the volume is a non-decreasing function of the radius, $V^*\geq V(r_2)$. Using this we can charge the weight
of the cut to the total volume. 

\begin{theorem}\label{gvy}
Let $ 0\leq r_1 < r_2$ be two radii, and suppose that $V(r_1)$ is greater than $0$. Then with probability at least
$3/4$ a random radii between $r_1$ and $r_2$ has a cut $C(r)$ such that
$C(r)\leq 4(\frac{\ln V(r_2) - \ln V(r_1)}{r_2-r_1})V(r)$.
\end{theorem}

The proofs of both these theorems can be easily inferred from the proofs of the corresponding versions in~\cite{GVY,VV}.
We outline them for completeness.

Let $u$ and $v$ be any two vertices and let $w\in S$ be such that $dist(u,S)=d_{uw}$. 
Then $dist(u,S)+d_{uv} = d_{uw}+d_{uv}\geq d_{vw}\geq dist(v,S)$, so that $\frac{d_{uv}}{dist(v,S)-dist(u,S)}\geq 1$.

From the definition of $V(r)$ we see that it is a 
piecewise linear non-decreasing function of $r$. It is differentiable at all points except possibly at those values of $r$
at which a new vertex arises. Since $dV(r)/dr = \sum_{u\in S(r), v\not\in S(r)}w_{uv} d_{uv}/(dist(v,S)-dist(u,S))$, using
the above inequality we get the following.

\begin{eqnarray}
\frac{dV(r)}{dr} & \geq & C(r)\label{derivative} \\
\int_{r_1}^{r_2} C(r) dr & \leq & V(r_2)-V(r_1)\label{volintegral}
\end{eqnarray}

\begin{proof} (of Theorem~\ref{average})
Let $C_{av}$ be the average value of the cuts between $r_1$ and $r_2$. From equation (\ref{volintegral}) it is seen that
$C_{av} \leq \frac{V(r_2)-V(r_1)}{r_2-r_1}$. Since only $1/4$ fraction of the radii may exceed $4$ times the average,
the result follows.
\end{proof}

\begin{proof} (of Theorem~\ref{gvy})
Let $\eps = 4\frac{\ln V(r_2) - \ln V(r_1)}{r_2-r_1}$, and let $B$ be the set of the radii between $r_1$ and $r_2$
that have a cut such that $C(r) > \eps V(r)$. Let $\mu(B)$ denote the measure of $B$. Then from equation
(\ref{derivative}), for $B$ we have $dV(r)/dr > \eps V(r)$. Integrating over $B$ we get

\begin{eqnarray*}
\int_B \frac{dV(r)}{V(r)} & > & \int_B \eps dr\\
\hbox{ i.e. } \ln \frac{V_{B_M}}{V_{B_m}} & > & \eps \mu(B).
\end{eqnarray*}

$V_{B_M}$ and $V_{B_m}$ denote the maximum and minimum values of the volumes in $B$. If $\mu(B) > \frac{r_2-r_1}{4}$
then,

\begin{eqnarray*}
V_{B_M} & > & V_{B_m} \exp(\eps \mu(B))\\
        & \geq & V(r_1) \exp(\ln V(r_2) - \ln V(r_1))\\
        & > & V(r_2).
\end{eqnarray*}

\noindent This is a contradiction as the volume is a non decreasing function. This implies that 
$\mu(B) \leq \frac{r_2 -r_1}{4}$ which proves the theorem.
\end{proof}

Finding a radius that corresponds to a small cut can be done deterministically in $O(n)$ steps~\cite{VV}.

\section{LP Rounding}
Solve the LP for {\it BMC} and consider the metric $(V,d)$ on $V$.
Give an initial volume of $\frac{V^*}{2k}$ to each source and sink, where $V^*=\sum_{e\in E}w_e d_e$ is the total volume.
The algorithm is as follows.

\begin{enumerate}
\item Choose a source-sink pair $s_i$ and $t_i$. Choose a radius $0<r_i<\frac{1}{4}$, such that 
$C(s_i,r_i) \leq (16\ln 4k) V(s_i,r_i)$ and $C(t_i,r_i) \leq (16\ln 4k) V(t_i,r_i)$ (we show below that such 
a common radius exists and can be found in polynomial time).
\item Put the vertices of $B(s_i,r_i)$ in $A$ and those of $B(t_i,r_i)$ in $B$.
\item Delete $B(s_i,r_i)$ and $B(t_i,r_i)$ from $G$, and repeat the procedure till no more sinks and sources are left.
\end{enumerate}

The correctness of the algorithm follows from the following easy claims.

\begin{claim}
At each step, a feasible radius exists.
\end{claim}

Let $r_2=1/4$ and $r_1 = 0$. Then $V(r_1) = V(0) = \frac{V^*}{2k}$ (since we start from a source or
a sink vertex which is given an initial volume of $\frac{V^*}{2k}$), and $V(r_2)$ is at most the total volume, $2V^*$.
Then by Theorem~\ref{gvy}, at least $3/4$ of the radii in $B(s_i,1/4)$ (respectively $B(t_i,1/4)$) are such that
$C(s_i,r) \leq 16(\ln 4k)V(s_i,r)$ (respectively $C(t_i,r)\leq 16(\ln 4k)V(t_i,r)$). Consequently, at least half of the 
radii are simultaneously suitable for both the regions of expansion. Thus a feasible radius exists.
Note that we search for this radius deterministically in linear time.

\begin{claim}
The graph $G_i$ at the $i$-th step is symmetric for all $i$.
\end{claim}

Since the graph is initially symmetric, this is true when $i$ is one. Suppose at the $i$-th step $G_i$ is symmetric.
Let $(s_i, t_i)$ be the source and sink pair from which we grow a region to radius $r_i$. Since the distance between
a source and the corresponding sink is at least one, (by inequality (\ref{everypath})) $B(s_i,r_i)$ and $B(t_i,r_i)$
cannot contain a source-sink pair.
Since by equations (\ref{symmetry}), a source/sink in $B(s_i,r_i)$ has its corresponding sink/source in $B(t_i,r_i)$,
$G_{i+1}$ is symmetric, proving the claim.

\begin{claim}
The algorithm returns a feasible solution to {\it BMC} within a factor of $O(\log k)$.
\end{claim}

We note that the graph left at each stage is symmetric. Also, at any stage a source/sink in $A$ has its corresponding
sink/source in $B$ and {\it vice versa}. This implies that the partition $A$ and $B$ of $V$ obtained is feasible
for {\it BMC}.
For the approximation factor, we note that the final value of the cut (denoted by $Cut$)
is no more than the sum of the values of the two cuts at each step. Hence, 
\begin{eqnarray*}
2V^* & \geq & V(s_1,r_1) + V(t_1,r_1)) + \ldots  \\
    & \geq & \frac{1}{16\ln 4k}[(C(s_1,r_1)+C(t_1,r_1))+\ldots ]\\
    & \geq & \frac{Cut}{16\ln 4k}
\end{eqnarray*}
This proves that the LP based rounding algorithm gives an approximation factor of $O(\log k)$.

\section{SDP Rounding}
We solve the SDP and obtain vectors $x_u$ for every vertex $u$. The distances $d_{uv}=|x_u-x_v|^2$, from the SDP 
give a metric space on $V$. Such metric spaces, where the square of the Euclidean distances form 
a metric are called as {\sf NEG}-metric spaces or $\ell_2^2$-metric spaces. An $\ell_2^2$ metric space on $n$ points
with all vectors of unit length is said to be spreading if the sum of the distances between points is $\Omega(n^2)$.
For unit $\ell_2^2$ spreading metric spaces there is a powerful structure theorem which we use.

\begin{theorem} [{\sf Arora, Rao, Vazirani}~\cite{ARV},{\sf Lee}~\cite{Lee}]\label{arv}
Let $(X,d)$ be an $n$-point $\ell_2^2$ metric space with $diam(X)\leq 1$ and
$\frac{1}{n^2}\sum_{x,y\in X}d(x,y)\geq \alpha>0$. Then there exist subsets $A,B\subseteq X$ with $|A|,|B|=\Omega(\alpha n)$
and $d(A,B)\geq 1/O(\sqrt{\log n})$, where the $O(.)$ notations hides some dependence on $\alpha$.
\end{theorem}

Let $M$ be the submetric of $V$ which has only the sources and sinks in $V$ as its vertices. Clearly, every symmetric 
subset of $M$ is a unit $\ell_2^2$ space. It turns out that it is also spreading.

\begin{lemma}\label{spreading}
Every symmetric subset of $M$ spreads.
\end{lemma}
\begin{proof}
Let $M'$ be a symmetric subset of $M$ and let (without loss of generality) $(s_1,t_1),(s_2,t_2),\ldots,(s_l,t_l)$ be its source-sink pairs.
Then the sum of the $\ell_2^2$ distances between pairs of points is at least 

\begin{eqnarray*}
& = & \frac{1}{2}\sum_i\sum_j( |x_{s_i}-x_{s_j}|^2 + |x_{t_i}-x_{t_j}|^2 + |x_{s_i}-x_{t_j}|^2 + |x_{t_i}-x_{s_j}|^2) \\
& = & \frac{1}{2}\sum_i\sum_j ( 2 |x_{s_i}-x_{s_j}|^2 + 2 |x_{s_i}+x_{s_j}|^2 )\qquad\hbox{ using antipodal constraints} \\
& = & \frac{1}{2}\sum_i\sum_j ( 4|x_{s_i}|^2 + 4|x_{s_j}|^2 )\\
& = & (2l)^2.
\end{eqnarray*}
Since $M'$ has $2l$ vertices, the lemma is proved.
\end{proof}

Using the antipodal constraints and the symmetry of $M$ we use the following theorem to obtain two antipodal subsets of $M$
that are separated by a distance $\Omega(\frac{1}{\sqrt{\log |M|}})$, each of size $\Omega(|M|)$.

\begin{theorem}[{\sf Agarwal-Charikar-Makarychev-Makarychev}\cite{ACMM},{\sf Lee}\cite{Lee}]\label{arvsymmetric}
Any symmetric unit $\ell_2^2$ representation with $2n$ points contains $\Delta$-separated w.r.t. the $\ell_2^2$ 
distance subsets $S$, and $T=-S$ of size $\Omega(n)$, where $\Delta=\Omega(1/\sqrt{\log n})$. Furthermore, there is a 
randomized polynomial-time algorithm for finding these subsets $S$, $T$.
\end{theorem}

The algorithm proceeds iteratively, with the invariant that $G$ is
symmetric. At the $i^{th}$ step, the graph is denoted by $G_i$ and the submetric on the sources and sinks by
$M_i$. The number of source-sink pairs in $M_i$ is denoted by $k_i$, so that there are $2k_i$ vertices in $M_i$.
A radius $r$ is called {\it good} for a set $S$ if either $C(S,r)\leq \frac{c}{\D}\frac{V^*}{\log 2k}$ or
$C(S,r)\leq c' \frac{\ln\ln 2k}{\D}V(S,r)$, where $c,c'$ are constants. The algorithm is as follows.

\begin{enumerate}
\item Using Theorem~\ref{arvsymmetric} on $M_i$, obtain two antipodal subsets $S_i$ and $T_i$ of size $\Omega(2k_i)$,
and separated by a distance $\Delta_i=\Omega(\frac{1}{\sqrt{\log 2k_i}})$.
\item Find a radius $r_i\leq\D_i/4$ which is {\it good} for both $S_i$ and $T_i$ simultaneously.
(We show below how to obtain it. In fact the proof shows that a random radius is good with constant probability).
\item Put the vertices of $B(s_i,r_i)$ in $A$ and those of $B(t_i,r_i)$ in $B$.
\item Delete $B(s_i,r_i)$ and $B(t_i,r_i)$ from $G_i$ to get the graph $G_{i+1}$, and repeat the procedure till 
no more sinks and sources are left.
\end{enumerate}

\section{Finding a Good Radius}
Let $\D_i$ denote the separation between the sets $S_i$ and $T_i$ obtained using Theorem~\ref{arvsymmetric} 
at the $i$-th step, and let $V_t = \frac{V^*}{\log 2k}$. The common {\it good} radius $r_i$, we find will
be at most $\D_i/4$, so that $B(S_i,r_i)$ and $B(T_i,r_i)$ are disjoint.

Consider $B(S_i,\frac{\D}{16})$ and $B(T_i,\frac{\D}{16})$. If the volume contained inside each is at most $V_t$,
then using Theorem~\ref{average} with $r_1=0$ and $r_2 = \D_i/16$, we see that at least $3/4$-th of the radii in $[0,\D_i/16]$
satisfy 

\begin{eqnarray*}
C(S_i,r)  & \leq & \frac{4}{\D_i/16}[V(\frac{\D_i}{16})-V(0)]\\
          & \leq & \frac{64}{\D_i}V_t
\end{eqnarray*}

\noindent (and similarly $C(T_i,r)\leq \frac{64}{\D_i}V_t$).
Consequently, at least half the radii in $[0,\D_i/16]$ are {\it simultaneously} good for both the regions of expansion. 

If the volume contained inside each is at least $V_t$, then using Theorem~\ref{gvy} with $r_1=\D_i/16$ and $r_2=\D_i/8$,
we see that at least $3/4$-th of the radii in $[\D_i/16,\D_i/8]$ satisfy 

\begin{eqnarray*}
C(S_i,r) & \leq & 4\frac{\ln V(\D_i/8) - \ln V(S_i,\D_i/16)}{\D_i/8-\D_i/16}V(r) \\
         & \leq & \frac{64}{\D_i}\ln(\frac{V^*}{V^*/\log 2k})V(r) \\
         &   =  & \frac{64}{\D_i}(\ln\log 2k)V(r)
\end{eqnarray*}

\noindent(and similarly $C(T_i,r)\leq \frac{64}{\D_i}\ln\log 2k V(r)$). Consequently, at least half
the radii in $[\D_i/16,\D_i/8]$ are {\it simultaneously} good for both the regions of expansion.

Now suppose (without loss of generality) that the volume inside $B(S_i,\D_i/16)$ is at least $V_t$ while that inside
$B(T_i,\D_i/16)$ is less than $V_t$. We have the following two cases depending on the volume of $B(T_i,\D_i/8)$.

If $V(T_i,\D_i/8)< V_t$, then let $r_1 = \D_i/16$ and $r_2=\D_i/8$. Theorem~\ref{gvy} applied to $B(S_i,\D_i/16)$
implies that with a probability at least $3/4$, a random radius in $[\D_i/16,\D_i/8]$ is such that
\begin{eqnarray*}
C(S_i,r) & \leq & 4 (\frac{\ln V(S_i\D_i/8) - \ln V(S_i, D_i/16)}{\D_i/8 - \D_i/16}) V(S_i,r) \\
         & \leq & (\frac{64}{\D_i} \ln \frac{V^*}{V^*/\log 2k}) V(S_i,r) \\
         & \leq & (\frac{64}{\D_i}\ln \log 2k )V(S_i,r).
\end{eqnarray*}
\noindent Theorem~\ref{average} applied to $B(T_i,\D_i,16)$ implies that with a probability at least $3/4$, a random
radius in $[D_i/16,D_i/8]$ is such that 
\begin{eqnarray*}
C(T_i,r) & \leq & \frac{4}{\D_i/8-\D_i/16}V^* \\
         &   =  & \frac{64}{\D_i} V^*
\end{eqnarray*}
\noindent Thus with probability at least $1/2$, a random radius in $[\D_i/16,\D_i/8]$ is simultaneously
good for both the regions of expansion.

If $V(T_i,\D_i/8) > V_t$, then let $r_1 = \D_i/8$ and $r_2=\D_i/4$. We apply Theorem~\ref{gvy} to both $B(S_i,\D_i/8)$
and $B(T_i,\D_i/8)$. Then with probability at least $3/4$ a random radius in $[\D_i/8,\D_i/4]$ is such that
\begin{eqnarray*}
C(S_i,r) & \leq & 4(\frac{\ln V(\D_i/4) - \ln V(\D_i/8)}{\D_i/4 - \D_i/8})V(S_i,r)\\
         & \leq & ( \frac{16}{\D_i}\ln\log 2k )V(S_i,r)
\end{eqnarray*}
\noindent Similarly, $C(T_i,r) \leq ( \frac{16}{\D_i}\ln\log 2k )V(T_i,r)$ for at least $3/4$-th of the radii in
$[\D_i/8,\D_i/4]$. Hence with a probability of at least $1/2$ a random radius in $[\D_i/8,\D_i/4]$ is simultaneously
good for both the regions of expansion.

Again, we can find the good radii deterministically in polynomial time.
Using an argument similar to the LP based algorithm, we see that the solution returned by this algorithm is feasible for
{\it BMC}. 
Since a constant fraction of the remaining vertices of $M$ are deleted at every step, there are $O(\log 2k)$ iterations.
Let $Cut$ be the value of the final cut obtained. This is at most the sum of the cut values obtained at each iteration.
For the approximation factor, we note that since there are $O(\log 2k)$ iterations, the total contribution of the cuts
charged to the total volume (i.e. by the application of Theorem~\ref{average}) is $O(\log 2k )\frac{1}{\Delta}\frac{V^*}{\log 2k}$
which is $O(\sqrt{\log k})V^*$. 

For the cuts obtained by the application of Theorem~\ref{gvy}, we note that since the volumes are deleted, their sum is 
at most $V^*$. 

\begin{eqnarray*}
V^* & \geq & \sum_i V(S_i,r_i) + \sum_j V(T_j,r_j) \qquad\hbox{ sums over volumes obtained using Theorem~\ref{gvy} }\\
    & \geq & \Omega(\frac{\D}{\ln\log 2k})[\sum_i C(S_i,r_i) + \sum_j C(T_j,r_j)] \\
    & \geq & \Omega(\frac{\D}{\log\log 2k})Cut \\
\end{eqnarray*}

Thus the approximation ratio is $O(\sqrt{\log k}\log\log k + \sqrt{\log k})= O(\sqrt{\log k}\log\log k)$. This improves
upon the LP based algorithm. 

\section{Integrality Gap of the SDP}
\def\O{\mathcal{O}}
\def\F{\mathcal{F}}
\def\V{\mathbf{V}}
\def\T{\mathbf{T}}
\def\W{\mathbf{W}}
In this section we show that the above SDP for BMC has an integrality gap of $\Omega(\log\log k)$. For this we use the
construction of Devanur, Khot, Saket and Vishnoi~\cite{KVDS}\footnote{All notations in this section follow~\cite{KVDS}}.
Let $\F_1,\F_2$ be the cubes $\{-1,1\}^N$ for some large prime $N$
and let $\O_1,\O_2,\ldots,\O_n$ denote the {\it orbits} as in~\cite{KVDS}. For $\mathbf{x} \in \F_1\times\F_2$, we say that
$\mathbf{x}$ and $\mathbf{-x}$ are complementary. Further, if two orbits $\O$ and $\O'$ are such that $\mathbf{x} \in \O$
{\it if and only if} $\mathbf{-x}\in\O'$, then we say that $\O$ and $\O'$ are complementary and denote $\O'$ by $-\O$.
Let $\sigma^i$ be a rotation operation~\cite{KVDS}. 

\begin{claim}\label{claim1}
If $\O$ is {\it nearly orthogonal}\cite{KVDS}, then so is $-\O$.
\end{claim}

\begin{claim}\label{claim2}
A {\it nearly orthogonal} orbit cannot contain a point and its complement.
\end{claim}

Claim (\ref{claim1}) follows directly from definition (3.1) of \cite{KVDS}. For claim (\ref{claim2}), note that
since $N$ is odd, the number of $1$'s in $\mathbf{x}$ and $\mathbf{-x}$ have different parities. Hence $\mathbf{-x}$
cannot be obtained by a rotation operation on the coordinates of $\mathbf{x}$.

Let $\O=\{\V_{\O,1},\V_{\O,2}\ldots,\V_{\O,N}\}$ denote any orbit, where $\V_{\O,1}$ is fixed arbitrarily, and
$\V_{\O,j} = \sigma^{j-1}(\V_{\O,1})$. Let $\V_{\O,j} = (\V_{\O,j}^x,\V_{\O,j}^y)$. Note that $\V_{\O,j}^x$ and
$\V_{\O,j}^y$ are in $\{-1,1\}^N$. Let $r,s$ and $t$ be as in~\cite{KVDS}. Since the $r$ used is even, the following holds.

\begin{claim}\label{claim3}
$\T_{\O,j}^x = (\frac{1}{\sqrt{N}}\V_{\O,j}^x)^{\otimes r} = \T_{-\O,j}^x $
\end{claim}

Let $\W_{\O,j}$, $1\leq j \leq N$ denote the vectors obtained by applying the Gram-Schmidt process on
$\T_{\O,j}$, $1\leq j\leq N$. From claim (\ref{claim3}), $\W_{\O,j} = \W_{-\O,j}$. Let
\begin{eqnarray*}
\V_\O  & = & (\frac{1}{\sqrt{N}}\sum_{j=1}^N y_j(\W_{\O,j}^x)^{\otimes 2s})^{\otimes t}
\end{eqnarray*}
Since $t$ is odd and the $y_j$'s in $\O$ and $-\O$ are complementary, we have the following claim.

\begin{claim}\label{claim4}
$\V_\O = - \V_{-\O}$
\end{claim}

Let $G$ be the multi-graph as in~\cite{KVDS} and let the source-sink pairs consist of orbits and their complements.
Then a feasible solution for {\it BMC} corresponds to a $(1/2,1/2)$ balanced cut. Using claim (\ref{claim4}), it is seen that
the vectors $\V_{\O_i}$ are feasible for the {\it BMC} SDP. Since every $(1/2,1/2)$ balanced cut is also a $(1/3,2/3)$ balanced
cut, using Theorem 2.3 of~\cite{KVDS}, we see that the {\it BMC} SDP has an integrality gap of $\Omega(\log\log k)$, where
$k=n/2$ is the number of source-sink pairs.

\appendix \label{multicutpath} 
\section{Multicut on Paths}

Let $P_n=v_n,e_n,v_{n-1},e_{n-1},\ldots,v_1,e_1,v_0$ be a path with a multicut instance defined on it.
Whenever we consider an induced multicut instance on a sub path
$P_i=v_i,e_i,v_{i-1},e_{i-1},\ldots,v_0$, the source-sink pairs considered shall only be those that have both their vertices
in $P_i$. We define two
operations on the highest numbered edge (the "leftmost" edge) of the path: $e^-_nP_{n-1}$ and $e^+_nP_{n-1}$.
$e^-_nP_{n-1}$ denotes the multicut instance on the sub path $v_{n-1},e_{n-1},\ldots,v_1,e_1,v_0$ obtained by
deleting the edge $e_n$ from $P_n$ and all source-sink pairs that have $v_n$ as one of its vertices from the 
source-sink list, and this is easily seen to be the instance on $P_{n-1}$. 
$e^+_nP_{n-1}$ denotes the multicut instance on the sub path $v_{n-1},e_{n-1},\ldots,v_1,e_1,v_0$
obtained by deleting the edge $e_n$ and replacing each source-sink pair $\{v_n,v_x\}$ by $\{v_{n-1},v_x\}$. If $e^*_i$
denotes either $e^+_i$ or $e^-_i$, then
$e^*_ne^*_{n-1}\ldots e^*_iP_{i-1}$ is defined recursively as the $e^*_{i}$ applied to the path
$e^*_ne^*_{n-1}\ldots e^*_{i+1}P_i$.
The instance $e^+_ne^+_{n-1}\ldots e^+_{j+1}e^-_jP_{j-1}$ is the same as the instance $P_{j-1}$.
This is easy to see as the two paths have the same underlying graphs. Also, since no end-points of the edges
$e_n,e_{n-1},\ldots,e_{j+1}$ is present in $P_{j-1}$, their source-sink pairs are also the same.

Let $OPT(P)$ be the optimal value of the multicut on a path $P$ and let $w_e$ denote the
weight of an edge $e$. Since an edge is either present or not present in the optimal solution to a multicut instance,
we have the following recursion.
\begin{eqnarray*}
OPT( e_i^+ e_{i-1}^+\ldots e_j^+P_{j-1} ) & = & w_{e_j}+OPT(P_{j-1})\hbox{ (if }v_j,v_{j-1}\hbox{\small form a source-sink pair in }e_i^+\ldots e_j^+P_{j-1}\hbox{)}\\
                                          & = &\hbox{ min }[ w_{e_j} + OPT(e_j^-P_{j-1}),\; OPT(e_j^+P_{j-1})] 
\end{eqnarray*}

We calculate $OPT(P_n)$ for the optimal. We can implement this recursion using a dynamic program.
\end{document}